\documentclass[sigconf,nonacm,natbib=false]{acmart}

\usepackage{cleveref}

\AtBeginDocument{%
  }

\RequirePackage[
  datamodel=acmdatamodel,
  style=acmnumeric,giveninits=true,maxbibnames=3
  ]{biblatex}

\begin{document}

\title{Comprehensive Plugin-Based Monitoring of \\Nexflow Workflow Executions}

\author{Sami Kharma}
\email{kharma@zib.de}
\orcid{0000-0003-0963-3771}
\affiliation{%
  \institution{Zuse Institute Berlin}
  \city{Berlin}
  \country{Germany}
}

\author{Tobias Wies}
\email{tobias.wies@tu-darmstadt.de}
\orcid{0009-0005-6485-3605}
\affiliation{%
  \institution{TU Darmstadt}
  \city{Darmstadt}
  \country{Germany}
}

\author{Florian Schintke}
\email{schintke@zib.de}
\orcid{0000-0003-4548-788X}
\affiliation{%
  \institution{Zuse Institute Berlin}
  \city{Berlin}
  \country{Germany}
}

\renewcommand{\shortauthors}{S. Kharma et al.}

\maketitle

\section{Background and Objectives}
Nextflow~\cite{nextflow} is a workflow management system commonly used in fields like bioinformatics~\cite{scenic} and earth observation~\cite{FORCE-on-Nextflow-2021}. It coordinates distributed data processing of various tools as an acyclic sequence
of tasks while using, containerization (e.g., Docker), orchestration (e.g., Kubernetes), or batch processing (e.g., SLURM). Monitoring such workflow executions can be challenging but aids performance analysis, debugging, and data provenance.

Besides Nexflow's basic built-in monitoring, the wf-commons tool~\cite{wfcommons} for creating wf-instances is widely regarded as the standard in the Nextflow community. The monitoring plugin we present provides a more detailed and flexible alternative compatible with wf-instances while removing the need for a custom Nextflow fork by using Nextflow's plug-in mechanism (version 21.10), optional direct .jar file changes of static artifacts without recompilation, and allows online monitoring during execution.

\section{Monitoring Methodology}

We implemented a Nextflow plug-in in Groovy\footnote{\url{https://github.com/cookiephone/nf-bigbrother}} that enables detailed workflow execution monitoring through callbacks and JVM reflections on Nextflow's internal state. It constructs the (partial) physical execution graph online at runtime.
For every task, it gathers detailed execution information such as start and endtime, execution method, container image, and the task's working directory, which can be used by other tools like schedulers for online processing of monitoring data. The plug-in itself uses this data to acquire additional context, i.e. by observing the shared filesystem in use to capture file metadata. It can be activated by enabling it in Nextflow's config file or with a command-line override. It provides valid and extended wf-instance JSON format.

Since Nextflow's available internal data or files do not expose detailed hardware information or allow access to it, we provide an optional patch to enable deeper node-level monitoring. This is done by injecting code or monitoring tools directly on the nodes a task is deployed to by Nextflow. This optional patch simply adds hooks to the wrapper script that Nextflow uses for task execution and thus allows to start external node-level monitoring tools to obtain CPU, RAM, network information, and IP addresses statically or resource usage as time-series, for example, using collectl.\footnote{\url{https://github.com/sharkcz/collectl}}

\section{Results}

\begin{figure}[t!]
\includegraphics[width=.8\linewidth]{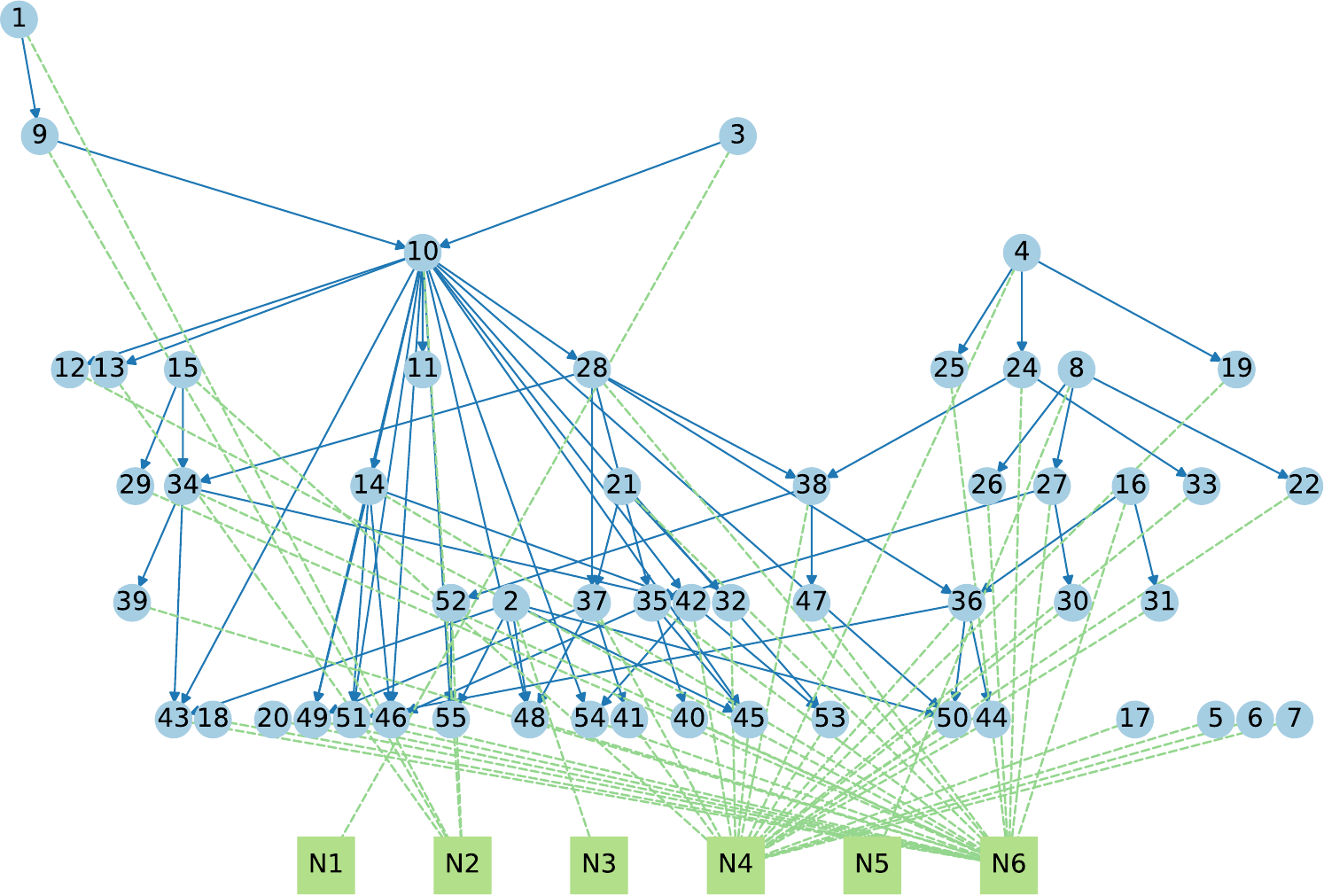}
\caption{Physical task graph (blue) and node (green) assignment of a partial rnaseq~\cite{rnaseq} run monitored with our plugin.\label{fig:rnaseq}}
\end{figure}

Using the plugin, data on real workflow executions is seamlessly collected without modifying the workflow. We monitored workflows from the nf-core repository~\cite{nf-core}. The assignment of physical tasks of a six-node execution of the rnaseq~\cite{rnaseq} workflow is shown in Fig.~\ref{fig:rnaseq}. The broadly expanded view on workflow executions, compared to the basic wf-instances monitoring, offers new opportunities for practical workflow analysis and optimization.

\section{Conclusion}
The presented plugin allows comprehensive tracing of Nextflow  executions without modifying the monitored workflow. The detailed data acquired aid insights into the workflows themselves, their performance and potential bottlenecks, but can also be leveraged to steer resource allocation and scheduling of workflows in ways classical HPC use-cases cannot.

\section*{Acknowledgements}
This work received funding from the German Research Foundation (DFG), CRC 1404: \emph{FONDA: Foundations of Workflows for Large-Scale Scientific Data Analysis}.

\printbibliography

@ARTICLE{nextflow,
  title    = "Nextflow enables reproducible computational workflows",
  author   = "Di Tommaso, Paolo and Chatzou, Maria and Floden, Evan W and
              Barja, Pablo Prieto and Palumbo, Emilio and Notredame, Cedric",
  journal  = "Nat Biotechnol",
  volume   =  35,
  number   =  4,
  pages    = "316--319",
  month    =  apr,
  year     =  2017,
  address  = "United States",
  language = "en"
}

@software{rnaseq,
  author       = {Harshil Patel and
                  Jonathan Manning and
                  Phil Ewels and
                  Maxime U Garcia and
                  Alexander Peltzer and
                  Rickard Hammarén and
                  Olga Botvinnik and
                  Adam Talbot and
                  Gregor Sturm and
                  nf-core bot and
                  Matthias Zepper and
                  Denis Moreno and
                  Pranathi Vemuri and
                  Mahesh Binzer-Panchal and
                  Ezra Greenberg and
                  silviamorins and
                  Lorena Pantano and
                  Robert Syme and
                  Gavin Kelly and
                  Lorenzo Sola and
                  Friederike Hanssen and
                  James A. Fellows Yates and
                  Gabriel Lichtenstein and
                  Jose Espinosa-Carrasco and
                  rfenouil and
                  Luke Zappia and
                  Chris Cheshire and
                  Edmund Miller and
                  marchoeppner and
                  Peng Zhou},
  title        = {nf-core/rnaseq: nf-core/rnaseq v3.19.0 - Tungsten
                   Turtle
                  },
  month        = jun,
  year         = 2025,
  publisher    = {Zenodo},
  version      = {3.19.0},
  doi          = {10.5281/zenodo.15631172},
  urlhidden          = {https://doi.org/10.5281/zenodo.15631172},
}

@article{nf-core,
  author    = {Philip A. Ewels and Alexander Peltzer and Sven Fillinger and Harshil Patel and Johannes Alneberg and Andreas Wilm and Maxime Ulysse Garcia and Paolo Di Tommaso and Sven Nahnsen},
  title     = {The nf-core framework for community-curated bioinformatics pipelines},
  journal   = {Nature Biotech.},
  volume    = {38},
  number    = {3},
  pages     = {276--278},
  year      = {2020},
  doi       = {10.1038/s41587-020-0439-x},
  urlhidden       = {https://doi.org/10.1038/s41587-020-0439-x},
  issn      = {1546-1696}
}

@article{wfcommons,
    title   = { {WfCommons: A Framework for Enabling Scientific Workflow Research and Development} },
    author  = {Coleman, Tain\~a and Casanova, Henri and Pottier, Lo\"ic and Kaushik, Manav and Deelman, Ewa and Ferreira da Silva, Rafael},
    journal = {FGCS},
    volume  = {128},
    number  = {},
    pages   = {16--27},
    doi     = {10.1016/j.future.2021.09.043},
    year    = {2022},
}

@ARTICLE{scenic,
  title    = "A scalable {SCENIC} workflow for single-cell gene regulatory
              network analysis",
  author   = "Van de Sande, Bram and Flerin, Christopher and Davie, Kristofer
              and De Waegeneer, Maxime and Hulselmans, Gert and Aibar, Sara and
              Seurinck, Ruth and Saelens, Wouter and Cannoodt, Robrecht and
              Rouchon, Quentin and Verbeiren, Toni and De Maeyer, Dries and
              Reumers, Joke and Saeys, Yvan and Aerts, Stein",
  journal  = "Nat Protoc",
  volume   =  15,
  number   =  7,
  pages    = "2247--2276",
  month    =  jun,
  year     =  2020,
  address  = "England",
  language = "en"
}

@inproceedings{FORCE-on-Nextflow-2021,
  author       = {Fabian Lehmann and
                  David Frantz and
                  S{\"{o}}ren Becker and
                  Ulf Leser and
                  Patrick Hostert},
  editorhidden       = {Gao Cong and
                  Maya Ramanath},
  title        = {{FORCE} on {Nextflow}: Scalable Analysis of Earth Observation Data on
                  Commodity Clusters},
  booktitlelong    = {{CIKM} 2021 Workshops},
  booktitlelong    = {Proceedings of the {CIKM} 2021 Workshops co-located with 30th {ACM}
                  International Conference on Information and Knowledge Management {(CIKM}
                  2021), Gold Coast, Queensland, Australia, November 1-5, 2021},
  serieshidden       = {{CEUR} Workshop Proceedings},
  volume       = {3052},
  publisher    = {CEUR-WS.org},
  year         = {2021},
  url          = {https://ceur-ws.org/Vol-3052/short12.pdf},
}

\vfill
\noindent This work was accepted as a poster at SCA/HPC Asia 2026.

\end{document}